\documentclass[fleqn,10pt]{wlscirep}

\usepackage[utf8]{inputenc}
\usepackage[T1]{fontenc}
\usepackage{placeins}

\title{Geographic characterization of railway systems}
\author[1,2,*]{Mark M. Dekker}
\affil[1]{Department of Information and Computing Sciences, Utrecht University, Princetonplein 5, 3584 CC Utrecht, The Netherlands}
\affil[2]{Centre for Complex Systems Studies, Utrecht University, Minnaertgebouw, Leuvenlaan 4, 3584 CE Utrecht, The Netherlands}
\affil[*]{m.m.dekker@uu.nl}

\begin{document}


\begin{abstract}
Railway systems provide pivotal support to modern societies, making their efficiency and robustness important to ensure. However, these systems are susceptible to disruptions and delays, leading to accumulating economic damage. The large spatial scale of delay spreading typically make it difficult to distinguish which regions will ultimately affected from an initial disruption, creating uncertainty for risk assessment. In this paper, we identify geographical structures that reflect how delay spreads through railway networks. We do so by proposing a graph-based, hybrid schedule and empirical-based model for delay propagation and apply spectral clustering. We apply the model to four European railway systems: the Netherlands, Germany, Switzerland and Italy. We characterize geographical structures in the railway systems of these countries and interpret these regions in terms of delay severity and how dynamically disconnected they are from the rest. The method also allows us to point out important differences between these countries' railway systems. For practitioners, this geographical characterization of railways provide natural boundaries for local decision-making structures and a first-order prioritization on which regions are at risk, given an initial disruption.
\end{abstract}

\flushbottom
\maketitle
\thispagestyle{empty}

\section{Introduction}
\label{sec:1}

Transport systems provide a core function to our society, moving passengers and goods around the globe to allow for global trade, business and leisure. Efficient infrastructure and mobility has been found important for economies to grow \cite{Wbcsd2009}. However, at severe economic costs, the efficiency of many such systems is regularly affected by perturbations and subsequent spread of disruptions. Examples can be found in trade networks being affected by epidemics \cite{Ivanov2020}, earthquakes \cite{Inoue2019} or cyclones \cite{Shughrue2020}, energy and internet problems due to power outages \cite{buldyrev2010}, railway disruptions due natural hazards \cite{dekker2021publ, dekker2019plos, bhatia2015, Ludvigsen2014} and many more. The economic costs of such events, also on smaller scales, motivate scholars and practitioners to investigate the robustness and resilience of these systems to perturbations \cite{Pagani2019, Nogal2016, ouyang2012}, and to understand and predict subsequent evolution of perturbations \cite{goverde2010, dekker2020}. This paper focuses on understanding the structure of the evolution of such perturbations in a subset of transport systems: railway systems. 

Railway systems involve the on-time geographical transport of passengers or goods, utilizing resources (\textit{assets}, like physical trains or crew) according to a predefined schedule, from, via and towards nodes (e.g., stations) in a network, along certain routes or tracks (edges). While the on-time dynamics are described in a pre-defined schedule, of interest here are the temporal deviations from the schedule, which are referred to as \textit{delays}, calculated as the executed time minus the predefined time --- in situations with no delay at all, all resources run on time and no delay is present. Hence, no perturbation or associated dynamics is present. What sets apart railway systems from many other transport systems, is their dependence on a detailed pre-determined system, in some countries even down to the particular assets used per activity \cite{dekker2020}. This attribute, in combination with usually high usage of existing capacity commonly causes perturbations that are an interesting (and necessary) topic of study. In the remainder of this paper, when we talk about `dynamics' of railway systems, we actually refer to the spreading and change of \textit{delays} rather than the scheduled movement of the assets --- note that the latter is usually a well-optimized plan that contains, by construction, little interacting elements (although the scheme itself may be prone to cascading-like phenomena that in turn affect the delay dynamics \cite{dekker2020}).

The research on transport delays can be split into multiple parts. A first part focuses on understanding the origin of initial perturbations, involving research on natural disasters as mentioned above, but also on smaller-scale fluctuations of activity delays that may give rise to systemic delay generation \cite{wen2019, dekker2021publ, Fleurquin2013, kecman2015b}. Oftentimes, these studies are case-specific, due to the heterogeneous (and often external) nature of such initial perturbations. Another line of transport delay research focuses on prediction of delays: the simulation of how a currently delayed situation will evolve to a (future) situation. Many kinds of approaches to this problem exist: micro simulation tools including a high level of detail on each edge \cite{middelkoop2006, nash2004}, larger-scale stochastic and analytic models both on the planning and realization side \cite{goverde2010, schobel2012, Fleurquin2013, monechi2018, Dekker2021Belgium}, and in more recent years, machine learning or data-driven studies are contributing to the field \cite{dekker2019plos, kecman2015a, markovic2015, Oneto2018}. A third branch of research involves the understanding of delay dynamics on a more structural level: characterizing the nature and (semi-)universal `laws' without the specific aim of prediction. For example, finding macroscopic nation-wide patterns in railway delay \cite{dekker2019plos}, characterizing mechanisms of how trains pass delay onto other assets \cite{monechi2018}, how the interaction between crew, rolling stock and line planning can lead to delay cascades \cite{dekker2020}.

This paper contributes to the third branch of transport (or railway) delay literature and focuses on the large spatial scale that is usually associated with these systems. We aim to identify geographic regions that act as subsystems, partially independent in terms of delays, and these clusters are reinterpreted in terms of their role in the global system, which add to the understanding of the general structure of these often nation-wide systems. The identification of such regions contributes to the overall understanding of the dynamics of delays, providing answers to questions like `where is delay generally generated, propagated through and attracted?' and `what regions are dynamically near-isolated and may be treated as such?'. These questions are useful for practitioners to not only help in prediction methods, but also for more strategic decisions on infrastructure planning and updates of future schemes.

Identifying substructures in complex systems such as railways is commonly done using clustering algorithms, with applications ranging from physics to ecology \cite{Dam2020}, climate \cite{tantet2015} and even epidemics \cite{Scarpino2019}. The wide variety of applications comes with a multitude of clustering methodologies. Graph-based clustering is often based on random-walks and modularity-optimization principles \cite{newman2006}, like the famous Louvain clustering method \cite{blondel2008}, or on spectral properties of core graph matrices \cite{Luxburg2007}. A famous data-based clustering method is $K$-means \cite{MacQueen1967}, while many other methods exist. It is important to note the difference between identifying substructures in the network topology alone, and doing so by accounting for spatial structures in dynamical processes happening on top of the network topology \cite{Dekker2021temp}. In this paper, we focus on the latter, using spectral clustering. The main reason for choosing spectral clustering as opposed to any modularity-optimization tool is that spectral modes and the shape of the eigenspectrum reveal more than just the detection of communities (as we see later in Fig.~3 of this paper). But indeed, there are important advantages of modularity optimization, as well, such as the automatic optimization of the amount of detected clusters.

Also in transportation literature, clustering techniques have been applied, e.g., to assist real-time management, operations and decision making \cite{Chen2014, Yang2017, cerreto2018, Kadir2018}. Other papers apply clustering tools on statistical variables to identify general states in the system \cite{Xia2012, Lin2019, dekker2019plos}. Even though compartmentalizing the geographical system based on observed or simulated delay patterns is less common, it is not new, e.g. concerning the identification of communities in cargo ships \cite{Kaluza2010}, assessing topological properties in the Swiss railway network \cite{Erath2009} or quantifying resilience in the Indian railways \cite{bhatia2015}. However, combining data on the dynamics of the (delay of the) system, with data on the infrastructure has rarely been done in railway literature. This, and the subsequent interpretation of those found communities, is the goal of this paper. It is important to emphasize that mere (topological) clustering of only the infrastructure, neglecting all other dynamical and operational information, would provide clusters that have less of an operational meaning: delay does not diffuse equally along directions that are topologically equivalent. The relation between more general diffusion dynamics among subregions in a networks has been investigated theoretically before \cite{Siudem2019}, but to our knowledge, this has never been applied to characterize dynamical structures in railway systems.

The paper is structured as follows. The model and the associated clustering method is discussed and applied to a fictitious transport network in section~\ref{sec:2}. In section~\ref{sec:3}, we apply the methods to real data of four European railway systems and show the results of the model and clustering. We interpret the clusters in terms of connectivity and role in the country's delay in section~\ref{sec:4}. We end with several conclusive remarks in section~\ref{sec:5}.
\section{Methods}
\label{sec:2}

In this section, we build a graph-based model for delay dynamics based on infrastructure and empirical train delays. We start by separating the underlying infrastructure from the delay dynamics that happens on top. The underlying infrastructure, consisting of \textit{nodes}, being stations or departure/arrival locations, and \textit{edges} as the railway tracks between them, is assumed to remain invariant in this analysis --- different from, for example network perturbation analyses or node-failure transport problems \cite{bhatia2015, buldyrev2010, Pescaroli2016}. An important factor in delay propagation is spatial heterogeneity: e.g., edges with a higher frequency of trains, fewer parallel tracks and re-routing options, are more prone to propagate or amplify delay than other edges. Rather than purely looking at the topology of the underlying infrastructure network, it is this heterogeneity that defines the weights of the edges that largely impact the resulting spectral clustering. We note that there are also other types of heterogeneities that impact prediction or model quality, like temporal heterogeneities (e.g., due to parts of the day being much more busy than other parts) and heterogeneities among transport resources, that are not included in this model.

We distinguish \textit{endogenous} from \textit{exogenous} spatial heterogeneity. Endogenous heterogeneities are static properties that affect delay propagation. They are derived from what is built within the system's schedules, e.g., (planned) travel times and resource travel frequency. These endogenous factors are used to imply what portion of existing delay at a station is transported in each possible direction. Exogenous ones, in contrast, relate to dynamical factors affecting delays and are derived from delay statistics rather than from the static (non-delayed) schemes. These exogenous factors act as multipliers: if endogenous factors determine a certain delay to be transported in a certain direction, the exogenous factor amplifies or dampens this delay. Below, we build a model where we define both of these factors.

\subsection{Endogenous spatial heterogeneity}

Endogenous factors that impact delays come from within the system's internal predefined properties, like the schedule. The intuition is that edges with more traffic propagate delay more easily and busier routes involve more congestion. In particular, delays are carried along with trains in the direction of their trajectories and nodes are more easily affected by a neighboring node's delays if the link between them contains high-frequent traffic, a mechanism that is commonly used in delay propagation models \cite{monechi2018, Dekker2021Belgium}. Given a railway system's scheme, train departure and arrival times are known. From this, one can derive train frequencies $f_{ij}$ and average travel time $\tau_{ij}$ for every edge $ij$ between nodes $i$ and $j$. This leads to an \textit{endogenous heterogeneity factor} $\alpha$ that weighs every edge $ij$ in terms of how delay gets redistributed due to the scheme (illustrated in Fig.~\ref{fig:fig1}a):

\begin{eqnarray}
\label{eqn:alpha}
\alpha_{ij} =
\begin{cases}
\frac{f_{ij}}{\tau_{ij} \cdot \rho \cdot \sum_{j'}f_{ij'}} &\text{ if $i\neq j$}\\
0 &\text{ if $i=j$}
\end{cases}
\end{eqnarray}

where $\rho$ denotes an inertia parameter (explained below) and the summation over $j'$ refers to adjacent nodes to $i$. The inertia parameter relates to rate (in inverse units of time) of refreshing a node's delay. In other words, it is inversely related to the time it takes before all current delay is diffused away through all outgoing edges from this node: a higher $\rho$ means lower $\alpha$, i.e., slow (endogenous) diffusion of delay, meaning that the system is dynamically very fragmented, and vice versa. Although we realize that this mechanism, too, can be heterogeneous across space (which implies that $\rho$ may be variable over $i$ and $j$), we simplify by making $\rho$ a constant: this allows us to interpret $\alpha$ fully as a schedule-based quantity, rather than impacted by the infrastructure of stations.

\subsection{Exogenous spatial heterogeneity}

\begin{figure}[!h]
    \centering
    \includegraphics[width=1\textwidth]{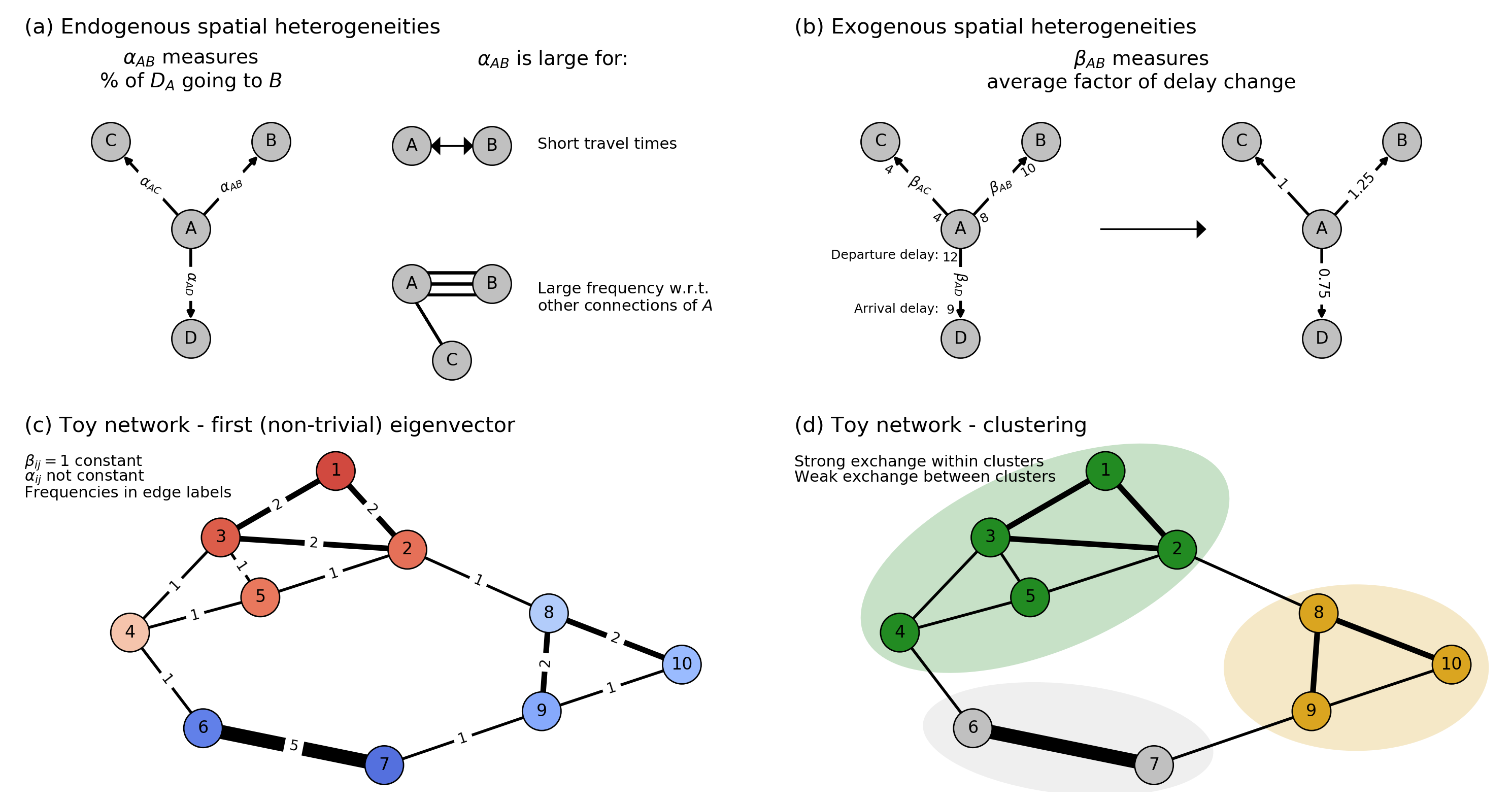}
    \caption{\textbf{Panel (a):} Illustration of the endogenous spatial heterogeneity $\alpha$.
    \textbf{Panel (b):} Illustration of the exogenous spatial heterogeneity $\beta$.
    \textbf{Panel (c):} First non-trivial eigenvector of an example network where we assumed $\beta$ constant, and $\alpha$ only variant because of changing transport frequencies (denoted in edge width and labels). Eigenvector coefficients are shown in node colors, from blue to red.
    \textbf{Panel (d):} Clustering of the example network, where $K=3$.}
    \label{fig:fig1}
\end{figure}

Besides endogenous factors, there are also exogenous factors that increase or decrease delays, depending on the area in the railway network. By `exogenous factors', we refer to those that are unrelated to the amount of traffic or other scheme-based factors. Examples of such factors are a large amount of traffic lights, fewer parallel tracks, speed limits, high frequencies of passengers (delaying boarding times), decreased vision and increased chances of infrastructure problems or trees falling on tracks. In our model, we derive these factors in an aggregated way from data by comparing delays of trains before and after crossing every edge. Specifically, we determine the \textit{exogenous heterogeneity factor} $\beta$ for each edge $ij$ as follows:

\begin{eqnarray}
\label{eqn:beta}
\beta_{ij} =
\begin{cases}
\frac{\langle D_{arr}^j \rangle}{\langle D_{dep}^i\rangle} &\text{ if $i\neq j$}\\
0 &\text{ if $i=j$}
\end{cases}
\end{eqnarray}

where $\langle D_{arr}^j\rangle$ and $\langle D_{dep}^i\rangle$ denote the average arriving and departing delays at node $j$ and $i$, respectively. In other words, it quantifies how much delay changes along this edge: $\beta_{ij} <1$ indicates that delays are, on average, decreased, while $\beta_{ij} >1$ indicates an average increase of delays when passing through edge $ij$. (We take the averages in both the numerator and the denominator in Eqn.~\ref{eqn:beta} to prevent near-zero arrival or departure times from strongly altering the $\beta$ value of the edge.) Examples and the intuition of $\beta$ are illustrated in Fig.~\ref{fig:fig1}b.

\subsection{Model}

The multiplications of the parameters $\alpha$ and $\beta$ result in weights of edges that reflect how delay propagates through a railway network. This can be viewed as a delay propagation model, although we never use it to actually simulate delays --- we use it to identify communities. The intuition of this model is as follows. Consider two stations I and II, connected by edge $ij$. Given an initial delay of $D_0$ at station I, the scheme predicts that a fraction $\alpha_{ij}$ of $D_0$ is propagated towards II (the rest, $(1-\alpha_{ij})\cdot D_0$ either remains at I or travels towards other stations). During the propagation of $\alpha_{ij} \cdot D_0$ towards II, the delay is increased or decreased on edge $e$ by a factor $\beta_{ij}$. In other words, the delay arriving at II in the next time step is $\alpha_{ij} \beta_{ij} D_0$. We generalize this reasoning by constructing the matrix $M$ by element-wise multiplication of the factors $\alpha$ and $\beta$:

\begin{eqnarray}
\label{eq:m}
M_{ij} = 
\begin{cases}
\alpha_{ij} \beta_{ij} &\text{ if $i\neq j$}\\
1 - \sum_k \alpha_{ik} \beta_{ik} &\text{ if $i=j$}
\end{cases}
\end{eqnarray}

As we are mainly interested in the dynamics and direction of delay propagation, rather than the absolute values of delay, we assume our model to conserve delay. This requires the matrix $M$ to become row-stochastic, which is realized by defining the diagonal elements as in Eqn.~\ref{eq:m}. From this equation, it becomes clear that $\rho$ in Eqn.~1 modulates the ratio between off-diagonal and diagonal elements of $M$: the one condition is that $\rho$ is taken such that $\sum_x \alpha_{ix} \beta_{ix}\leq 1$, because the diagonal elements of $M$ are not allowed to become negative (as delay is treated here as a purely positive quantity). We refer to the matrix $M$ as the \textit{dynamical} matrix, which, given an delay vector $\vec{D}(t)$ at time $t$, gives the delay vector at the next time step $t+\Delta t$ through the following multiplication:

\begin{eqnarray}
\label{eqn:model}
\vec{D}(t+\Delta t) = \vec{D}(t) + \Delta t \cdot M \cdot \vec{D}(t)
\end{eqnarray}

where $\Delta t$ refers to a time step, which is arbitrary and not of concern for the research in this paper. Both $\alpha$ and $\beta$ are dimensionless and we only use $M$ to find out \textit{structural} properties of the railway systems rather than for explicit delay modeling --- i.e., even though Eqn.~4 illustrates an interpretation of $M$, we do not use this equation.

\subsection{Properties of $M$}

Many important model properties can be derived analytically from the row-stochastic matrix $M$, and we focus our research to the properties of this matrix. (Note that $M$ contains averaged, time invariant, entries through the definitions of $\alpha$ and $\beta$ and therefore does not depend on time.) Any delay simulation done using $M$ eventually ends up with a homogeneous delay vector $\vec{D}_i(t) = \vec{c}$ at all nodes $i$, where $\vec{c}_i := c = \frac{\vec{D}(t_0)}{N}$ with $N$ the total amount of nodes, because, given enough time, delay is spread in all corners of the graph (assuming it being connected). The question of how delay spreads in the \textit{transient} is of interest to us, i.e., before this trivial homogeneous solution. One way of identifying structural properties of a graph-based model is to look at the spectral modes. The intuition behind this is to find geographical delay patterns $\mathcal{D}$ that are persistent, i.e. slow to diffuse. Such patterns reveal natural boundaries that delays might not easily cross and geographical divisions into regions where delay is easily exchanged. In 2.5, we discuss how we use these spectral modes to define clusters. Mathematically, the problem of finding spectral modes $\mathcal{D}$ is defined in the eigenproblem of matrix $M$: 


\begin{eqnarray}
\label{eqn:eigen}
M\cdot \mathcal{D} = \lambda\mathcal{D}
\end{eqnarray}

with eigenvectors $\mathcal{D}$ and eigenvalues $\lambda$. The closer $\lambda$ is to 1, the slower the delay decay in a (relative) delay distribution fixed by $\mathcal{D}$ (this follows directly from Eqn.~\ref{eqn:eigen}) --- i.e., the more persistent is the geographical delay pattern $\mathcal{D}$. In particular, there is one solution with $\lambda = 1$ such that $M\cdot \mathcal{D} = 1\cdot \mathcal{D}$, because $M$ is row-stochastic. This is the aforementioned homogeneous solution (i.e., constant everywhere), and corresponds to the first eigenvector. This does not provide any insight in the dynamical structure of the system, and hence will be referred to as the `trivial' eigenvector. The attribute of these eigenvectors being persistent, points to dynamical connections among nodes with approximately equal coefficients in an eigenvector. In other words, the coefficients of the eigenvectors can be used to find clusters of nodes that are have a dynamical connection, as we follow up on in the next section.

\subsection{Clustering}
\label{sec:2_clus}

As mentioned above, the clustering is based on the coefficients of the eigenvectors: nodes with roughly equal coefficients are dynamically connected (i.e., easily exchange delay among themselves) and, intuitively, should therefore be grouped in one cluster. A single eigenvector provides interesting information already, which is why we propose a spectral approach in this paper. Combining multiple eigenvectors, however, allows us to do the actual clustering. Let us assume that we are searching for $K$ clusters (the question of choosing $K$ is addressed later). A common method of finding these clusters is by creating a $K$-dimensional embedding based on the first $K-1$ non-trivial eigenvectors (as the first eigenvector is constant and does not add any information), and applying a $K$-means algorithm to this space \cite{Dam2020, Luxburg2007, Shi2000, Ng2002}. Indeed, this results in $K$ clusters that are based on the respective differences between eigenvector coefficients. There are many more clustering methods, with various advantages and disadvantages. The advantage of $K$-means clustering is that it is one of the most well-known and intuitive Euclidean distance-based clustering methods. The disadvantage of $K$-means is that many implementations of the algorithm are not deterministic, and that $K$ is not automatically defined.

We derive an appropriate value of $K$ based on the so-called `eigengap heuristic' \cite{Luxburg2007}, which is based on the eigenvalue spectrum. High (near-1) eigenvalues correspond to relatively persistent -- and thus to us important -- eigenvectors $\mathcal{d}$. Sudden `jumps' in the eigenspectrum therefore point to a group of more important eigenvectors (those with higher $\lambda$) and the rest of the spectrum (those with lower $\lambda$) and can thus be used to distinguish which eigenvectors are therefore of interest. Assuming an equal amount of clusters one can distinguish with this set of eigenvectors (although the trivial eigenvector is not useful in this analysis), the largest `eigengap' in the eigenspectrum defines $K$. Having a maximum amount of desired clusters (in the remained of this paper, we use $K_{max}=15$), results in:

\begin{eqnarray}
K = \text{max} _{i=2} ^{15} \{\lambda_{i-1}-\lambda_{i}\}
\end{eqnarray}

Summarized: we start with the eigendecomposition (Eqn.~\ref{eqn:eigen}), then from the eigenvalue spectrum we determine $K$, we continue by constructing the $K-1$-dimensional embedding with the first $K-1$ non-trivial eigenvectors (the minus-1 stems from excluding the trivial eigenvector), and apply $K$-means to the coefficients in this space. This results in $K$ clusters.

\subsection{Toy example}
\label{sec:2_toy}

We illustrate the model and associated clustering in a fictitious transport system consisting of 10 nodes and 14 edges in Fig.~\ref{fig:fig1}c-d. Trains go from node to node in a networked manner and their (bidirectional) frequencies are denoted in numbers on each edge, creating geographical differences in $\alpha$. The exogenous heterogeneity factor $\beta$ is assumed to be constant: $\beta = \beta_0=1$. The resulting $M$ matrix gives the first non-trivial eigenvector as displayed in colors in the nodes in Fig.~\ref{fig:fig1}c. By eye already, one can distinguish the red colored nodes (coefficients $>0$) from the the blue colored nodes (coefficients $<0$), which also makes sense dynamically: even though connections between nodes 4 and 6, and between 2 and 8 exist, they are much weaker than the interconnections between nodes 1-5, and subsequently do not bring the coefficients (i.e., of 4/6 and 2/8) of the first eigenvector close together. It turns out that for this system, $K=3$. The resulting clustering is found in Fig.~\ref{fig:fig1}d, confirming our observations by eye on the coefficient separation in Fig.~\ref{fig:fig1}c: the algorithm groups the interconnected region of 1-5. It also distinguishes nodes 6 and 7 from 8-10, as a result from the strong connection between 6 and 7. While this is only a fictitious toy network, in the following sections, we apply the same algorithm to real and much larger transport networks.

\FloatBarrier

\section{Application to European railway systems}
\label{sec:3}

We apply the algorithm to data of nation-wide railway systems of four European countries: the Netherlands, Italy, Germany and Switzerland. We have chosen these systems based on their relative comparability: railways in the United States, for example, have a strong emphasis on cargo transport (in contrast to European railways, having more emphasis on passengers) affecting frequency and regularity of the schemes, and even various topological aspects. Another example is the Chinese railway system, differing from European systems in terms of scale: having fewer stations (per unit area) and much longer travel times. Hence, for the illustration of the methods in this paper, we focus on four railway systems that are relatively comparable, but still have smaller cross-differences. We start this section by elaborating on the data itself and topological properties of these systems, after which we present the spectral results of the $M$ matrix for every of these four systems. We end this chapter with showing the resulting clusters.

\subsection{Data}

We utilize operational data from the Dutch, Italian, German and Swiss railway systems, including data on infrastructure, schedules (used to determine the values of $\alpha_{ij}$ for every edge $ij$) and realized delay data (used to determine the values of $\beta_{ij}$). Details on the source and cleaning of the data can be found in SI A. The data contains departure and arrival times, locations and their delays, per unique train number, along with infrastructure information on stations, their connections and longitude-latitude data. We obtain daily average variables like frequencies and travel times by looking at periods of 16 to 31 days (depending on the country, see SI A). Frequency is computed as average number of trains per hour by summing the total amount and dividing by 24. A minimum value of 1 minute travel time (in hours) is taken: i.e., lower average travel times are approximated to 1 minute. We use an inertia parameter $\rho=1000$ for all countries. In Tab.~\ref{tab:countries}, an overview of several static properties of the railway systems is given.

\begin{table}[!h]
    \centering
    \begin{tabular}{l|cccc}
        \hline
        Variable & Netherlands & Germany & Switzerland & Italy\\
        \hline
        \# connected nodes & 658 & 5815 & 1346 & 2201\\
        \# edges & 1438 & 15764 & 3827 & 6683\\
        Average degree & 2.26 & 2.90 & 2.56 & 3.26\\
        Average betweenness & 0.048 & 0.0052 & 0.0095 & 0.013\\
        Average daily \# train activities per node & 173 & 20.54 & 102.35 & 22.47\\
        Average daily \# unique service lines per node & 10.8 & 2.2 & 10.3 & 2.8\\
        Average $\alpha\cdot\rho$* & 13.01 & 4.13 & 8.34 & 2.43\\
        Average $\beta$* & 1.09 & 0.96 & 0.89 & 1.02 \\
        \hline
    \end{tabular}
    \caption{General overview of the four countries assessed in this paper. Node numbers are determined after removal of nodes that are not connected to the giant component. *Here, we show the average \textit{non-diagonal} $\alpha$ and $\beta$ values.}
    \label{tab:countries}
\end{table}

In Tab.~\ref{tab:countries}, we can see several differences across the four systems. When interpreting these numbers, it is good to emphasize that the distance between nodes and the level of detail varies across the four data sets: for example, in the Netherlands, the data is more detailed the level of passenger stations: there are sensors near the tracks (also outside of stations) that log whether trains are passing by. This level of resolution is higher for the Dutch case than in the other sets, affecting static properties of the system like reducing average travel time between edges or average degree, for example. (This resolution difference will not affect cross-system comparisons in later figures.) But even when taking this note into account, we conclude that the German railway system is clearly the largest, in terms of both nodes and edges. The low average degree and high average betweenness in the Netherlands can be explained by the fact that it includes many degree-2 nodes sequential on a line (by construction, but partly also due to the high resolution of the data), rather than having larger hubs that are interconnected.

\begin{figure}[!h]
    \centering
    \includegraphics[width=1\textwidth]{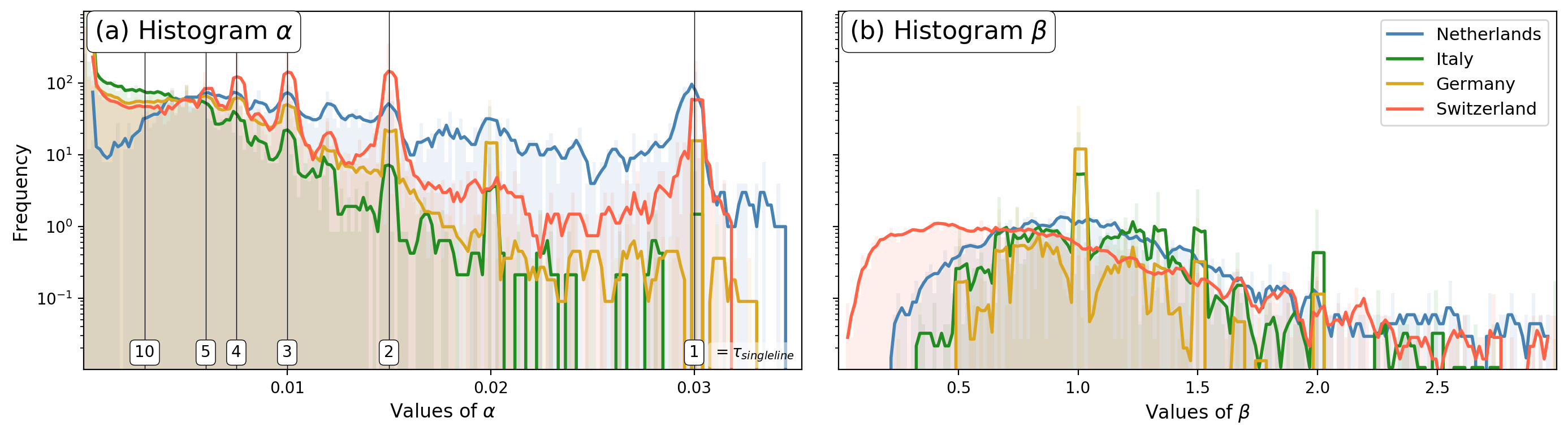}
    \caption{Normalized histograms of occurrence of $\alpha$ and $\beta$ values in the four analyzed countries, excluding diagonal elements of the $\alpha$ and $\beta$ matrices. In panel (a), the vertical lines indicate $\alpha$ values corresponding to integer travel times at edges where the frequency is 2 (i.e., one bidirectional service line). These, with potentially small variations in the frequency, dominate the signal in $\alpha$.}
    \label{fig:fig2}
\end{figure}

From a dynamical perspective, the values about service lines and train activities are more relevant. Service lines are defined as single-direction trips from a starting station to an ending station (commonly at the other side of the country), crossing various in-between stations. These lines are denoted with a unique number and are important advectors of delay \cite{dekker2020}, which is why their structure is important for railway dynamics. The smaller systems (Switzerland and the Netherlands) clearly dominate with respectively 10.3 and 10.8 unique lines per node, in comparison to Germany and Italy, having values of 2.2 and 2.8 respectively. This reflects the dynamically denser nature of the Dutch and Swiss railway systems, and is confirmed in the higher number of daily train activities per node (102-173 for the Netherlands and Switzerland as opposed to 20-22 for Italy and Germany).

The average $\alpha$ and $\beta$ values also have an important meaning. They are proxies for the density of trains --- modulated by frequency and travel time --- and statistical delay increases (on edges), respectively. A relatively high $\alpha\cdot \rho$ in the Netherlands and Switzerland depicts relatively high train density between stations and long tracks of degree-2 nodes to prevent scheduled fragmentation of delay spread (the $\rho$ factor, although constant across these countries is added to prevent bias compared to potential future work). The $\beta$ values are like a railway edge-performance metric: the lower, the less delay is increased along its edges. Interestingly, the dense, highly utilized Swiss railway network performs best when looking at these numbers, having an average $\beta$ of 0.89. (Note that for a full performance comparison among these railway systems, one should statistically correct for a number of factors like delay changes \textit{within nodes}, which is not included here.) The histograms of $\alpha$ and $\beta$ of the four countries are shown in Fig.~\ref{fig:fig2}a and b, respectively. In several countries, $\alpha$ values corresponding to average travel times of whole integers on tracks where only 1 line is traveling -- which makes sense because $\alpha$ reflects exactly such schedule-based biases in how delay spreads. This is most notable in Switzerland and the Netherlands, where we expect such short tracks (of 1, 2, 3, 4 or 5 minutes travel time) to exist, because of the high density of these railway systems. The Italian system has many more lower values of $\alpha$ than the other countries (with the Dutch system having the fewest), corresponding to either tracks connected to nodes with trains in many (other) directions, or tracks with large average travel times. In panel (b), the distributions of $\beta$ are wrapped around 1, corresponding to no significant average change in delay on these tracks. Furthermore, we see that Switzerland and the Netherlands have more $\beta$ values lower than 1 in comparison to Italy and the Germany -- the latter two mainly having values of $\beta$ close to 1.

\subsection{Spectral results}

\begin{figure}[!h]
    \centering
    \includegraphics[width=1\textwidth]{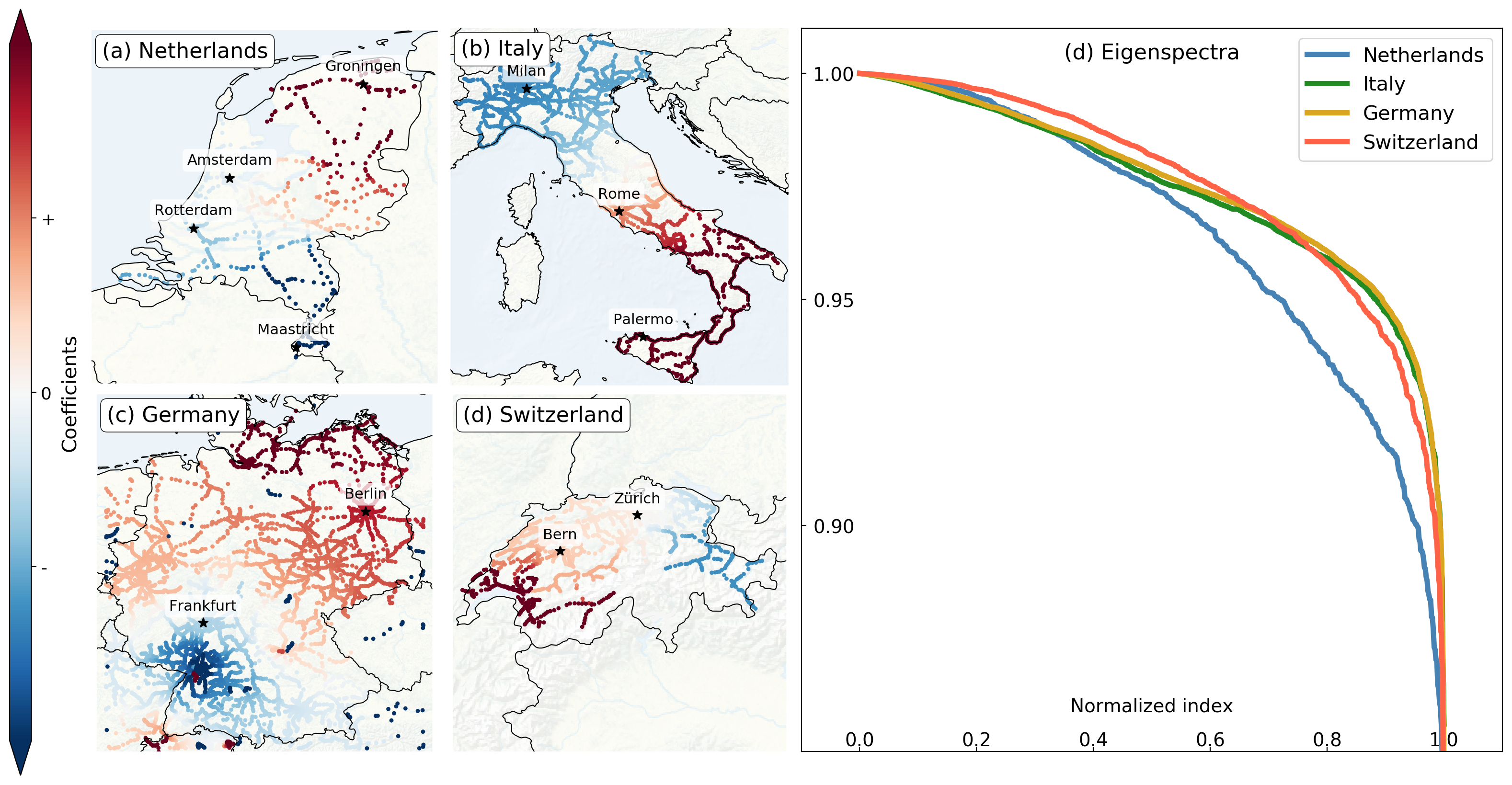}
    \caption{\textbf{Panel (a)-(d):} First non-trivial eigenvectors of the $M$ matrix, for the four European railway systems: (a) the Netherlands, (b) Italy, (c) Germany and (d) Switzerland. As absolute values of the coefficients are not relevant (only their relative magnitude to other ones in the same country are relevant), the colorbar and coefficients are shown normalized. Several important cities are annotated by black stars.
    \textbf{Panel (e):} Eigenspectra of the four countries, with normalized horizontal axis.}
    \label{fig:fig3}
\end{figure}

From the values of $\alpha$ and $\beta$, we determine the matrix $M$ using Eqn.~\ref{eq:m}. Fig.~\ref{fig:fig3} shows the first non-trivial (i.e., second) eigenvectors of the four countries in panels (a)-(d), and the associated eigenspectra in panel (e) (with a normalized horizontal scale).

The coefficients of the first non-trivial eigenvector in all four countries show a dipole-like structure. In the Netherlands (Fig.~\ref{fig:fig3}a), the coefficients depict a north-south gradient, mainly highlighting a northern region (near Groningen) and the very south-east (near Maastricht) from the rest of the country, with a better connected center in between, including cities like Rotterdam and Amsterdam. The Italian coefficients (Fig.~\ref{fig:fig3}b) also show a north-south gradient, but more smooth --- not highlighting near-disconnected part from the rest of the weighted network. Germany (Fig.~\ref{fig:fig3}c), shows a clear separation of the area south of Frankfurt (including cities like Karlsruhe and Stuttgart) in blue. Apparently, this area may have persistent delays that are not easily exchange with the rest of the country. The Swiss graph (Fig.~\ref{fig:fig3}d) shows, like in the Dutch and Italian case, a clear geographic gradient: from east to west, highlighting the south including the cities of Geneva, Lausanne and Sion on the west, and a more rural area in the east. The city of Z\"urich is somewhat in between. 

Figure~\ref{fig:fig3}d shows the eigenspectra of the four countries, with a normalized horizontal axis to compare the spectra independent of network size. The eigenvalues are relatively high. The slow decrease of the eigenspectra indicates that these are not strongly disconnected clusters. The relatively faster decrease of the Dutch eigenspectrum reveals that in the weighted network spun by $M$, the delay changes in Dutch railway system is more strongly dominated by the first few components, as opposed to others. This reflects a that the first $K$ eigenvector are better suited for compartmentalizing the Dutch railways into subregions, than they are for other countries. (We do note, however, that the relative levels of the eigenvalues are (as mentioned before) partly modulated by the inertia parameter $\rho$. We have now taken $\rho$ constant for all countries.)

The (simple) dipolic structure of these eigenvectors and the fact that there are many high-value eigenvalues in the spectrum both point to the need of multiple eigenvecotrs to obtain a more refined view of dynamical structures in these networks.

\subsection{Clustering results}

\begin{figure}[!h]
    \centering
    \includegraphics[width=1\textwidth]{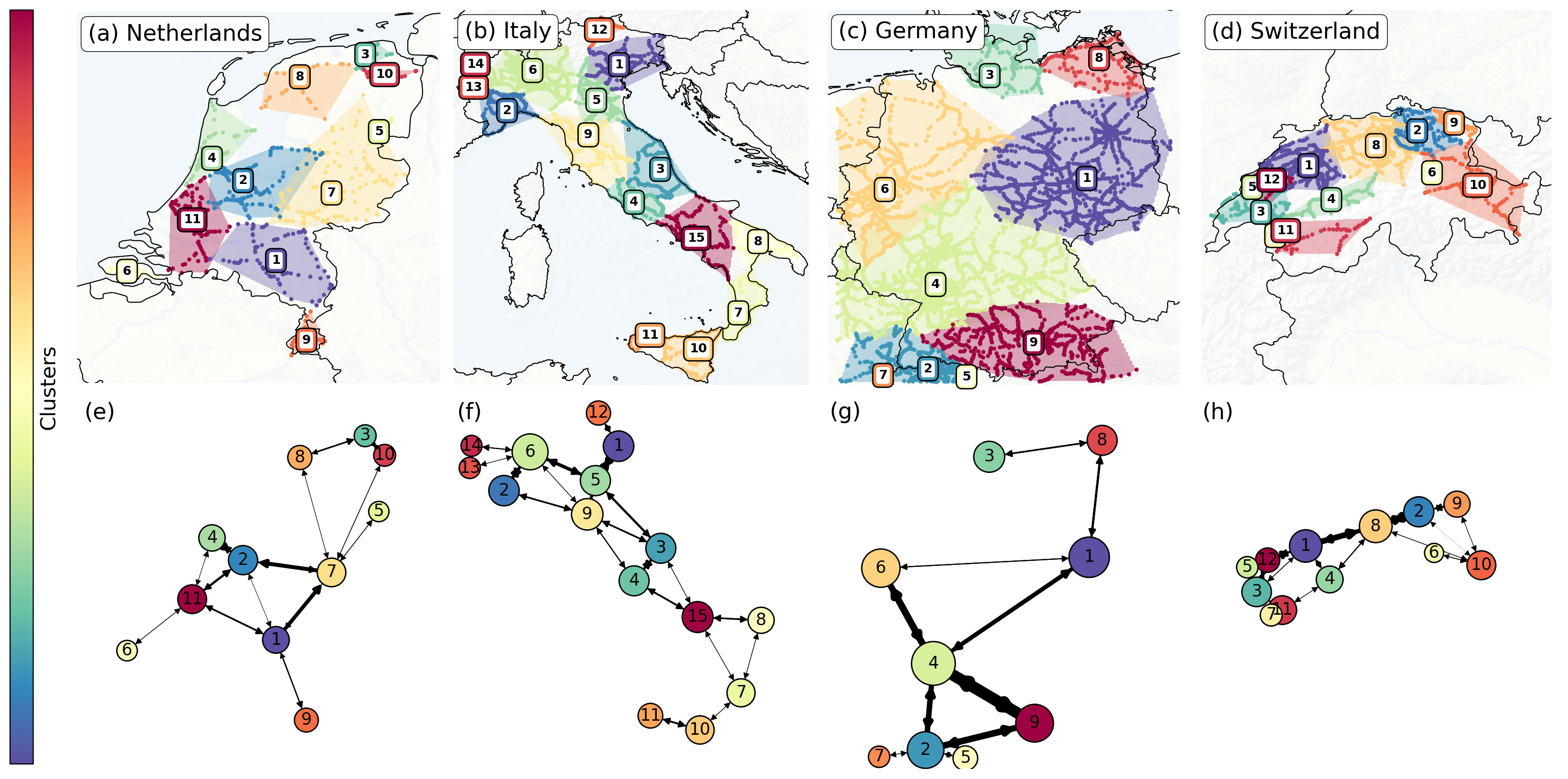}
    \caption{\textbf{Panel (a)-(d):} Clusters in the four European railway systems as found using $K$-means on the embedding built by the first $K-1$ non-trivial eigenvectors of the matrix $M$ (where $K$ is found using the eigengap heuristic). In particular, $K$ equals respectively 11, 15, 9 and 12 for these four countries.
    \textbf{Panel (e)-(h):} Abstracted versions of upper panels, reflecting the clusters and their connections. Arrows indicate connectivity between clusters in terms of $\alpha$ and $\beta$ --- the elements of $M$. The arrow width depicts the total sum of elements of $M$ directed from one cluster to another. Node size reflect amount of railway stations in each cluster.}
    \label{fig:fig4}
\end{figure}

Using the eigengap heuristic and the eigenspectra in Fig.~\ref{fig:fig3}d, we determine $K$ --- the number of clusters to search for, and the dimension of the embedding (see Sec.~\ref{sec:2_clus}. Subsequently, we apply the $K$-means clustering algorithm to the eigenspace and find the clusters shown in Fig.~\ref{fig:fig4}, with the geographic locations of the clusters in panels (a)-(d) and abstracted networks (including delay exchange in the arrow widths) in panels (e)-(h). 

In general, a total of 11, 15, 9 and 12 clusters are found in the railway systems of the Netherlands, Italy, Germany and Switzerland, respectively. By nature of the planar (2D-) structure of railway networks, the clusters are geographical intraconnected regions. In the Netherlands (Fig.~\ref{fig:fig4}a), the country is cut up in roughly equally-sized clusters, with exceptions of smaller clusters (3, 5, 6 and 10) at the edges of the country (e.g. in the rural areas of Zeeland in the south-west or Groningen in the north-east) --- probably as a result of service lines being less frequent and more disconnected from the center of the country in these areas. Easily spotted are the central cluster 2, including the major cities of Amsterdam and Utrecht, and cluster 11, including most of the so-called `Randstad', the most urbanised and transport-heavy area of the Netherlands. The clustering in Italy (Fig.~\ref{fig:fig4}b) also subdivides the country into more-or-less equal areas, with some exceptions of small clusters in the north (clusters 12, 13 and 14). Most major Italian cities are surrounded in a separate cluster: Rome (cluster 4), Milan (cluster 6), Genoa (cluster 2), Napels (cluster 15) and Venice (cluster 1). The island of Sicily is only weakly connected to the rest via a train ferry at the city of Messina, and the island is further subdivided into two other clusters. Germany (Fig.~\ref{fig:fig4}c), in contrast, is subdivided in only 9 clusters (although being the largest railway system among these four) that are not equal in size: clusters 1, 4, 6 and 9 add up to than 80\% of the country, including several foreign tracks. Cluster 6 almost perfectly coincides with the German federal states (`Bundesl\"ander') of Nordrhein-Westfalen, Brehmen and Niedersachsen, while cluster 1 covers most central-eastern states (these Bundesl\"ande are also used in operations by the dominant railway company, Deutsche Bahn). Several much smaller clusters can be found in the south: clusters 5 and 7, even being foreign (non-German) clusters and hence not of importance here. The large size of the clusters reflect that railway transport on the tracks are generally well distributed, travel long distances, and are less regional. The Swiss partitioning (Fig.~\ref{fig:fig4}d) results in rather small, intricately connected clusters (as is also visible in the abstracted graph in Fig.~\ref{fig:fig4}h). Several major urban areas can be recognized like the Geneva-Lausanne area (cluster 3) and the area around Bern (cluster 1), while the area around Z\"urich is divided into two clusters (cluster 2 and 8).

In the abstracted plots (panels (e)-(h)), the arrows and their widths provide information on the general flow of delay, proxied by the elements of $M$: thick arrows mean more (and larger) elements of $M$ between these clusters, indicating stronger pathways of delay propagation as dictated by the $\alpha$ and $\beta$ values, while small arrows mean the opposite. The Netherlands has a relatively well connected dynamical core in its center: exchanging most delay among the four clusters 1, 2, 7 and 11, containing several major transport corridors for both passengers and freight. The Italian plot indicates that most exchange is in the north, between clusters 1, 2, 5 and 6, reflecting a highly touristic and urbanized area, possibly leading to well-connected railway operations, as well as a strong connection between 3 and 4. The German clusters are mostly connected in the west and south: forming a strong bond between 2, 4, 6 and 9, reflecting some international transport and core lines from Cologne and Frankfurt to Stuttgart and M\"unchen. The core connections of the clusters in Switzerland are mostly along the west and northern border, connecting cities of Geneva, Lausanne, Bern and Z\"urich.

\section{Interpretation and relation to daily operations}
\label{sec:4}

The clusters reflect regions where delay is expected to to be propagated within the cluster, and less so towards outside of the cluster, incorporating both exogenous and endogenous effects. This section is devoted to interpret and understand the relevance of these clusters in terms of the delay and daily operations. We start by defining two metrics that allow such interpretation and apply them to the clusters found in Fig.~\ref{fig:fig4}. Subsequently, we compare the four countries and their clusters in terms of these metrics.

\subsection{Metrics to interpret clusters}

To find the operational and dynamical meaning of the clusters found in Fig.~\ref{fig:fig4}, we first define a few relevant quantities that we later combine into two cluster-characterizing metrics. We attribute any train's delay to its \textit{departure} location. The first quantity is the total delay $D_{\text{total}}$ from stations in the cluster. Second, we determine the internal delays $D_{\text{int}}$: delays of trains departing and arriving within the cluster. Third, exported delays $D_{\text{exp}}$ of trains departing from within the cluster, but arriving in another, are computed. And fourth, the imported delays $D_{\text{imp}}$ of trains departing from another cluster, arriving inside the analyzed cluster. The following relation holds: $D_{\text{total}} = D_{\text{int}}+D_{\text{exp}}$. With these quantities, we compute two metrics which allow for easy interpretation of the clusters: (1) their fraction of the country-wide delays, measured by the \textit{cluster severity} and (2) their dynamical (dis-)connectedness to other clusters, proxied by the \textit{cluster independence}.

Cluster severity $S(n)$ for any cluster $n$ is the ratio of $D_{\text{total}}$ of cluster $n$ to the average $D_{\text{total}}$ over all clusters:
\begin{eqnarray}
\label{eqn:sn}
S(n) = \frac{D_{total}(n)}{\frac{1}{N}\sum_i^N D_{total}(i)}
\end{eqnarray}
with $N$ being the total amount of clusters. So, if $S(n) >1$, the cluster covers an above-average part of the delays in the country, and vice versa. This does not incorporate cluster size, meaning that if all stations cover equal amounts of delay, larger clusters immediately have larger values of $S(n)$. We have chosen to not account for cluster size to make $S(n)$ a property of the cluster as a whole, make it better interpretable and relate to the practical use of the metric: small clusters might otherwise attain very high values of $S(n)$ while in practice not that dominant in delay.

Cluster independence $I(n)$ for any cluster $n$ is defined as the ratio between delays exchanged internally in the cluster, w.r.t. the delays exchanged with other clusters:
\begin{eqnarray}
\label{eq:In}
I(n) = \frac{D_{int}(n)}{\sqrt{G(n)}\cdot (D_{imp}(n)+D_{exp}(n))}
\end{eqnarray}
where $G(n)$ is the number of nodes in cluster $n$. This factor is included to account for the cluster size when comparing $D_{int}(n)$ to $D_{imp}(n)+D_{exp}(n)$ -- otherwise, from a basic geometric point of view, $I(n)$ automatically decreases with cluster size. This can also be shown when perceiving cluster $n$ as a simple square grid with a total amount of stations $G$. The stations at the border grow in numbers with $4\sqrt{G}-4$, which are the stations exchanging delay with other clusters, i.e. they modulate $D_{imp}(n)+D_{exp}(n)$. The number of internal stations (total minus border), proxying $D_{int}(n)$, equals $G-(4\sqrt{G}-4) = G-4\sqrt{G}+4$. In other words, the fraction $\frac{D_{int}(n)}{D_{imp}(n)+D_{exp}(n)}$ is proportional (for large $G$) to $\sqrt{G}$, a conclusion that also holds when assuming circular cluster grids. This is why we divide by $\sqrt{G}$ to avoid the metric $I(n)$ to be biased by the cluster size. Note that we do account for the cluster size bias in $I(n)$, but not in $S(n)$. This might seem inconsistent. However, $S(n)$ is already normalized by the average value of total delays in the clusters, while $I(n)$ is not a normalized value. The quotient of internal delays versus delay exchange (i.e., $I(n)\cdot \sqrt{G(n)}$) will vary greatly and therefore, adding $\sqrt{G(n)}$ as a normalization factor benefits the interpretation of $I(n)$ to be this quotient relative to the cluster size.


The metrics $S(n)$ and $I(n)$ both surround values of 1: values lower than 1 depict clusters that cover a less-than-average delays and are well-connected, respectively, and values higher than 1 reflect clusters with more-than-average delays and that are less connected. This allows to split the $I(n)-S(n)$ plane into four parts, that can be used to interpret the clusters. Four cluster categories can be distinguished, as displayed in Tab.~\ref{tab:clustercat}: named for easier reference Type A, Type B, Type C and Type D clusters.

\begin{table}[!h]
    \centering
    \begin{tabular}{l|cc}
        \hline
         & $S(n)<1$ & $S(n)>1$\\
        \hline
        $I(n)>1$ & \textbf{Type A}: & \textbf{Type B}: \\
        & weakly connected, small amounts of delay & weakly connected, large amounts of delay\\
        $I(n)<1$ & \textbf{Type C}: & \textbf{Type D}: \\
        & well connected, small amounts of delay & well-connected, large amounts of delay \\
        \hline
    \end{tabular}
    \caption{Cluster categories in the $I(n)-S(n)$ plane.}
    \label{tab:clustercat}
\end{table}

\subsection{Metric results}

\begin{figure}[!h]
    \centering
    \includegraphics[width=1\textwidth]{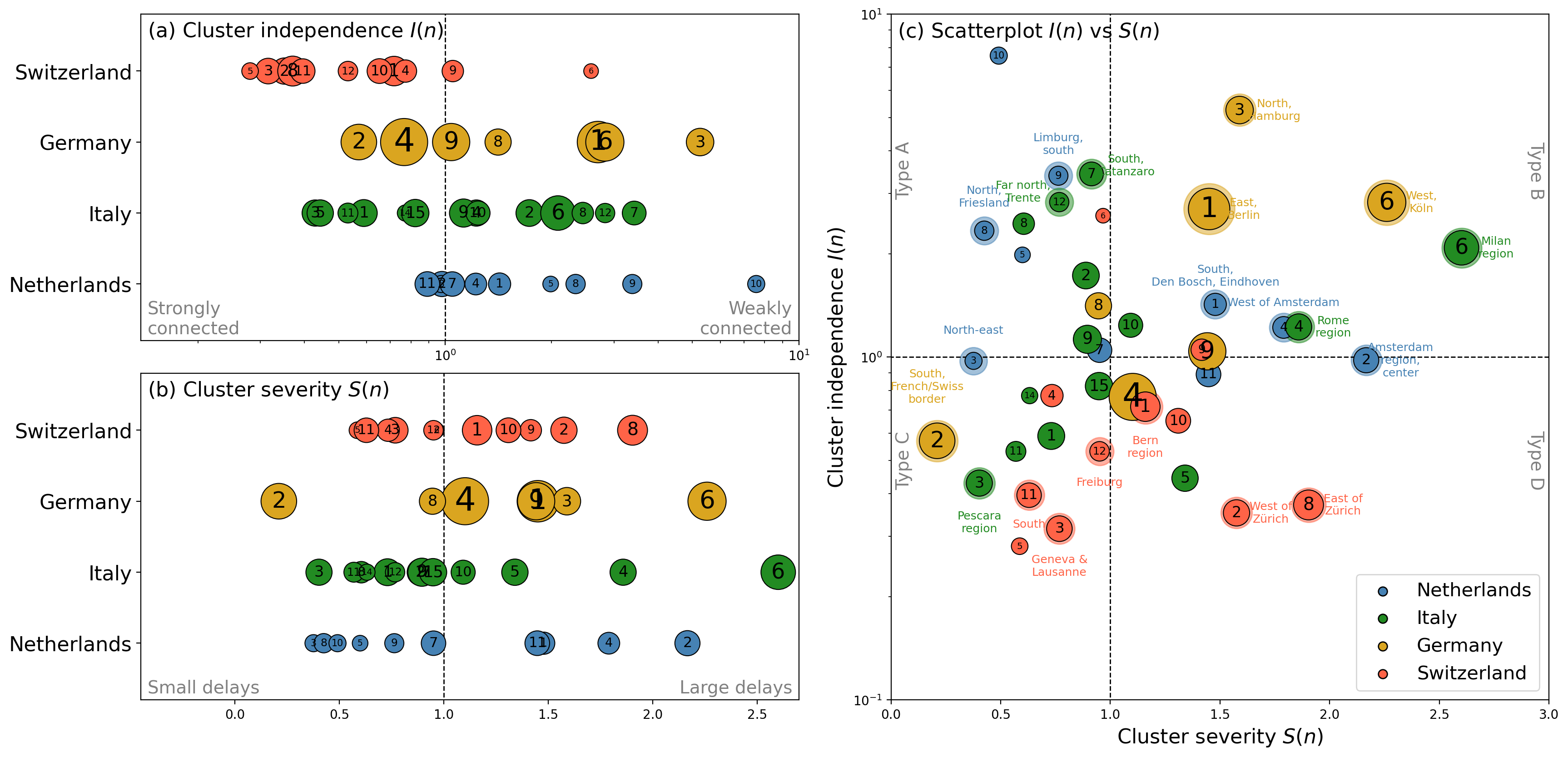}
    \caption{Cluster independence $I(n)$ and cluster severity $S(n)$ of the clusters across the four analyzed countries. \textbf{Panels (a)-(b)}: $I(n)$ and $S(n)$, respectively, on the horizontal and countries on the vertical. \textbf{Panel (c)}: Scatterplot with $I(n)$ versus $S(n)$. Several notable clusters are annotated in text and their dots are emphasized. Dot size reflects the number of stations inside the cluster. Clusters with only a very small amount of data (e.g. due to being across the border, involving only international trains) are not included in this figure. Numbers in the clusters refer to numbers in Fig.~\ref{fig:fig4}.}
    \label{fig:fig5}
\end{figure}

The values of $I(n)$ and $S(n)$ for all clusters $n$ across the four countries (in colors) are plotted in Fig.~\ref{fig:fig5} (a few clusters with too little data to estimate $I(n)$ and $S(n)$ well are left out). The numbers in each circle refer to the numbers in Fig.~\ref{fig:fig4}.

The panels (a) and (b) allow for both an in-country comparison of the clusters as well as a cross-country comparison of the spread of the $I(n)$ and $S(n)$ values. In panel (a), the immediate observation is the relatively low $I(n)$ values in Switzerland in contrast to relatively high values in the Netherlands, with Germany and Italy approximately in between. Apparently, many clusters in the Swiss system exchange more delays with other systems than they have internally. This may be a consequence of the fact that this relatively small country has many trains connecting clusters, rather than traveling in a separated space. This points to a potential source of vulnerability: delay can quickly spread across the country. Looking more closely to the cluster numbers within each country, we can find mostly urban and central areas having low $I(n)$ values per country, such as the area of Z\"urich (numbers 2, 8) in Switzerland, Pescara (3) in Italy and the regions near Amsterdam (2, 4) in the Netherlands. Higher $I(n)$ values are obtained by rural regions such as the south (9) and north (10) in the Netherlands and the south (7) and north (12) of Italy. An exception seems to be Germany, where the clusters are so large that they convey well-separated, but urban areas, such as the big central clusters (1, 6) and the area around Hamburg (3).

The $S(n)$ values in panel (b) are by definition spread around the value of $S(n)=1$, because $S(n)$ is normalized by its average (Eqn.\ref{eqn:sn}). Still, the spread provides valuable information on these countries. While the Swiss clusters are relatively well spread across the average, the German systems have a few clear outliers on both sides: the southern border (number 2) clearly having the smallest fraction of delays, whereas the area around K\"oln (6) has the largest fraction of delays, neither of both are interestingly the \textit{largest} cluster, which is number 4. Both Italy and the Netherlands show many clusters in the low-$S(n)$ domain, with a few outliers with high $S(n)$ values, mostly in the bigger cities -- Amsterdam (2, 4) and Eindhoven (1) in the Netherlands, and Rome (4) and Milan (6) in Italy.

In panel (c), the clusters are shown in the combined $I(n)-S(n)$ plane. The top-left quadrant (Type A clusters) mainly comprises Dutch and Italian rural clusters, such as the far north or south in these countries: the areas of Friesland and Limburg in the Netherlands and those around Catanzaro and Trente in Italy. The rural nature of clusters in this category is no surprise, as their periphery-located position in the network make them usually less connected with the rest (i.e., high $I(n)$) and less prone to delays (i.e., low $S(n)$). The Type B clusters ($I(n), S(n)>1$) are mainly found in Germany, with the busy, well-connected areas around Berlin, K\"oln and Hamburg. One other notable cluster is the urban Milan region in the north of Italy (cluster 6), having a mediocre value of $I(n)$, but an exceptional level of $S(n)$, reflecting a 2.5 times larger-than-average role in the total delay of the country. At the bottom-left we see the Type C clusters, containing mostly Swiss and Italian clusters that are quite well connected (like Pescara and Geneva), but play less of a role in the total delay. The Type D clusters (bottom-right), are mainly found in Switzerland, interpreted as well-connected clusters that also play a large role in the total delay. The Swiss Type D clusters are the relatively urban regions in this country around Z\"urich and Bern.

Summarized, comparing countries in the $I(n)-S(n)$ plane reveals that Switzerland generally has low values of $I(n)$, implicating the strong connectivity of this railway network. Germany, not really represented in the Type A and Type D categories, has a few large clusters in Type B. This may reflect that German delays are usually quite compartmental: large delays may arise but these large clusters mainly keep these delays within. The Italian and Dutch clusters show a large spread for values $S(n)<1$, and several clusters around $I(n)=1, S(n)>1$. There seem to be a few clusters in both of these countries that determine a disproportionally large amount of the total delay (high $S(n)$), but these clusters have an average connection with the rest of the country (unlike several German clusters). Overall, the patterns in Fig.~\ref{fig:fig5} seem to relate to operational characteristics, topological embedding in the general railway network and even urban differences across the clusters.

\section{Conclusion}
\label{sec:5}

To find geographical structures that have a dynamical meaning in railway systems, we developed a graph-based model and proposed a method to use spectral properties of this model to characterize the railway systems geographically.

In particular, combining both endogenous and exogenous spatial heterogeneities, all model information is encapsulated in the dynamical matrix $M$ whose spectral properties can be used to compartmentalize any transport system's geographical network into clusters. We analyze four European nation-wide railway systems: the Netherlands, Italy, Germany and Switzerland. Both infrastructural and operational data are used, revealing dynamical properties of the four countries (Tab.~\ref{tab:countries}) --- making an interesting comparison in itself --- and an optimal partition. The clusters vary in size and are connected in different ways to the nation-wide networks, resulting in the identification of core, central clusters and peripheric, near-disconnected clusters. The operational meaning of the clusters is expressed in two variables: cluster independence $I(n)$, reflecting the dynamical (dis-)connectedness of the cluster to the rest of the country, and cluster severity $S(n)$, reflecting the fraction of delay the cluster is responsible for. This leads to the classification of four cluster types, showing that high-$S(n)$ values are usually obtained by the more urban and dense clusters (e.g., high values relating to busy areas like the region around Milan, K\"oln and Amsterdam), and that $I(n)$ is partially distinguishes rural regions from central regions (e.g. far north and south of the Netherlands versus the regions of Geneva, Z\"urich and Pescara), but is partially also affected by how a country is operational handled. The analysis also shows that topological connectivity does not automatically determine dynamical connectivity, as shown in Fig.~\ref{fig:fig3}, e.g. resulting in rather central regions like K\"oln or Milan to have high $I(n)$ values.

Throughout the paper, various comparisons have been made between these four countries, revealing the Netherlands and Switzerland to be dynamically dense railway networks (i.e., having high utilization of their tracks), while Germany and Italy are much larger and sparser networks. Shorter steps between (logging) stations make the edges on average small in the Netherlands, with the longest edges in Italy. The clusters in Switzerland show low values of $I(n)$, depicting a strong interconnectedness among them. The Italian $S(n)$ is dominated by the Rome and Milan region, with almost all other regions having $S(n)<1$. The upper-right corner of the $I(n),S(n)$ plane is dominated by several German clusters near Hamburg, Berlin and K\"oln.

Railways, by schedule, have long-stretched service lines going from one side of the country to the other. Per definition, this means that perturbations quickly travel large parts of the country, as is also reflected in the high eigenvalues of $M$ (Fig.~\ref{fig:fig3}). The structure of the dynamics therefore does not favor partition much, and care should be taken into separating them too strongly in the interpretation of Figs.~\ref{fig:fig4} and \ref{fig:fig5}. Still, compartmentalization is an important topic in railways, in particular when it comes to defining regions to subdivide operations and dispatching. The proposed clustering should therefore be viewed more as a statistical average, or a composite of dynamical modes, rather than a full geographical segregation of delays. For future research, the connection and strength between the found clusters should be compared with delay propagation on a larger spatial scale: the relation between regional effects and nation-wide effects is still an important unresolved topic in transport literature. This work brings us one step closer to a solution.

For both scholars and railway practitioners, these results shape deeper understanding of how their railway systems work, how they differ from each other, and how regions with these systems each play a unique role. While the country-by-country statistics in Fig.~\ref{fig:fig2} and Tab.~\ref{tab:countries} add to system-wide insights, the clustering in Fig.~\ref{fig:fig4} may help practitioners to find boundaries when aiming to subdivide countries into smaller operational regions. Connections among the clusters may inform railway operators of statistically average directions of delay flows. And the characterization of the clusters by $I(n)$ and $S(n)$ in Fig.~\ref{fig:fig5} provides insights in how analogies between countries and regions can and cannot be made, and how different regions play different roles. For scholars, the construction and subsequent clustering of the $M$ matrix is a simple procedure and can be generalized to many other transport systems, beyond railways alone. While delay is a quantity strictly bound towards a predefined transport schedule, in theory many other dynamical variables on networks can be analyzed in the same manner. We believe that the presented methodology and results for the European countries contributes to deeper understanding of these systems, and we hope that the paper ignites more research in the relation between regional effects and nation-wide effects in transport systems.

\FloatBarrier




\section*{Acknowledgments}
The author thanks Debabrata Panja for his valuable comments on the manuscript.\\

\noindent \textbf{Funding:} This work is part of the research programme `Improving the resilience of railway systems' with project number 439.16.111, which is financed by the Dutch Research Council (NWO) and co-financed by Nederlandse Spoorwegen (NS) and ProRail. These organisations did not have a role in the study design or research conduction.\\

\noindent \textbf{Author contributions:} MMD conducted all research and text writing.\\

\noindent \textbf{Competing interests:} The authors declare no competing interests.\\

\noindent \textbf{Data and materials availability:} The analyses are done on publicly available data. For more information, see SI A.

\section*{Supplementary Materials}
SI A - Data sources and processing

\clearpage
\newpage
\appendix
\counterwithin{figure}{section}

\section{Data sources and processing}
\label{app:A}

\subsection{Data sources}
Operational data for the Netherlands is obtained from the Dutch infrastructure manager ProRail and covers the period July 2018 to June 2019. A selection of 16 days is made, based on delay severity (4 days in each of four categories; Green, Neutral, Red and Black, for more details see the Appendix or \cite{dekker2020} and \cite{dekker2019plos}), taken to statistically cover multiple situations. Network data on locations and links of the nodes and edges of the railway network, including delay time series can be found in an Open Science Framework related to previous work\cite{dekker2019plos}: \url{https://osf.io/tps4r/}. Data for the German and Italian railways are obtained from the supplementary material of \cite{monechi2018}, who received this data from the OpenDataCity (\url{http://www.opendatacity.de/}) and the ViaggiaTreno (\url{http://www.viaggiatreno.it/}) websites. The German data spans across March 2015, and the Italian data across March 2015 and April 2015. The Swiss railway data is obtained from the OpenTransportData website (\url{https://opentransportdata.swiss/}) over January 2018. The German data consists of several stations across the border in mainly the south and west, which are strongly connected to its lines.

\subsection{Data cleaning}
From the infrastructure data, the network topologies of these systems is determined. In all but the Dutch case, the data had to be cleaned for unrealistically long edges. These edges arise from trains travelling long distances between major cities without stopping in subsequent smaller stations, effectively adding direct edges between major cities that in practise are not direct edges. This cleaning is done by looking at the distribution of edge lengths.

Furthermore, it is important to note that there are disconnected (e.g. Sardinia in the Italian case) and non-operational components in some of the railway systems analysed in this paper. These parts are omitted - we focus on the giant component and where necessary, make all links bidirectional. Note that these problems only concern a small fraction of the links in these systems.

\subsection{Train types}
We focus on passenger trains only, and exclude freight trains in the analysis. One reason for this is that freight trains are (economically) privacy sensitive, meaning that it is difficult to get a complete dataset. Another reason is that delay is not always well-defined for such trains, as their routes and schedules are mostly separated from the main schedule. It should be noted that their contribution to delay is usually only minor, as, for example in the Netherlands, they cover only 5.7\% of all Dutch train kilometers in 2017 (numbers courtesy of the Dutch infrastructure manager ProRail). More details can be found in Dekker et al. (2019)\cite{dekker2019plos}, section 3.1.

\end{document}